\documentclass[journal]{IEEEtran}
\usepackage[T1]{fontenc}

\usepackage{cite}

\usepackage{subfigure}
   \usepackage[pdftex]{graphicx}
\usepackage{amsmath}
\usepackage{amssymb}
\usepackage{amsthm}
\usepackage{amsmath}
\usepackage{extarrows}

\usepackage[cmintegrals]{newtxmath}

\usepackage{array}
\newcommand{\tabincell}[2]{\begin{tabular}{@{}#1@{}}#2\end{tabular}}
\usepackage{booktabs}
\usepackage{bm}
\usepackage{lineno}
\usepackage{xcolor}

\begin{document}
%
\title{Deep Learning for Wireless Communications: An Emerging Interdisciplinary Paradigm}
%
%
%

\author{Linglong~Dai,
        Ruicheng~Jiao,
        Fumiyuki Adachi,
        H. Vincent Poor, 
        and Lajos Hanzo
\thanks{L. Dai and R. Jiao are with the Beijing National Research Center for Information Science and Technology (BNRist) as well as the Department of Electronic Engineering, Tsinghua University, 100084 Beijing, China (E-mails:
daill@tsinghua.edu.cn, jiaors16@mails.tsinghua.edu.cn).

F. Adachi is with the Research Organization of Electrical Communication, Tohoku University, Sendai, Japan (E-mail:
adachi@ecei.tohoku.ac.jp). 

H. V. Poor is with the Department of Electrical Engineering, Princeton
University, Princeton, 08544 New Jersey, USA (E-mail:
poor@princeton.edu). 

L. Hanzo is with the School of Electronics and Computer Science, University of Southampton, SO17 1BJ Southampton, UK (E-mail: lh@ecs.soton.ac.uk).}
 \thanks{This work was supported by the National Natural Science Foundation of
China for Outstanding Young Scholars (Grant No. 61722109), the Royal
Academy of Engineering through the UK-China Industry Academia Partnership
Programme Scheme (Grant No. UK-CIAPP$\backslash$49), and the U.S. National Science Foundation under Grant ECCS-1647198. L. Hanzo would also like to acknowledge the financial support of the European Research Council's Advanced Fellow Grant.}

\vspace{-4mm}
}
\maketitle



\begin{abstract}
Wireless communications are envisioned to bring about dramatic changes in the future, with a variety of emerging applications, such as virtual reality (VR), Internet of things (IoT), etc., becoming a reality. However, these compelling applications have imposed many new challenges, including unknown channel models, low-latency requirement in large-scale super-dense networks, etc. The amazing success of deep learning (DL) in various fields, particularly in computer science, has recently stimulated increasing interest in applying it to address those challenges. Hence, in this review, a pair of dominant methodologies of using DL for wireless communications are investigated. The first one is DL-based architecture design, which breaks the classical model-based block design rule of wireless communications in the past decades. The second one is DL-based algorithm design, which will be illustrated by several examples in a series of typical techniques conceived for 5G and beyond. Their principles, key features, and performance gains will be discussed. Furthermore, open problems and future research opportunities will also be pointed out, highlighting the interplay between DL and wireless communications. We expect that this review can stimulate more novel ideas and exciting contributions for intelligent wireless communications. 
\end{abstract}

\vspace*{-4mm}
\section{Introduction}
The way we live and work has been fundamentally changed by wireless communications over the past few decades. Thanks to the rapid development of the mobile Internet, billions of people all over the world are wirelessly connected through mobile phones for a wide range of daily activities, including social networking, information searching, etc. Historically, the prosperity of wireless communications has relied on its own model-based design paradigms, where accurate mathematical models and expert knowledge are required. However, the demanding requirements of the emerging applications, such as communicating under excessively complex scenarios with unknown channel models, low-latency requirement in large-scale super-dense networks \cite{FiveDis}, etc., are hard to be addressed by the traditional model-based wireless techniques. 
To be more specific, when the communication scenario cannot be readily described mathematically, 
either due to the excessively complex environment such as the underwater environment, or owing to  the non-linearity imposed by the unavoidable hardware impairments \cite{OFDM},
 the mathematical model based communication blocks, such as channel estimation, channel equalization, etc., will fail to accurately model the reality. Moreover, as emerging large-scale communication schemes (massive multiple-input multiple-output (MIMO) relying on a large number of antennas at the base station, massive Internet of things (IoT) scenarios connecting numerous users/devices, etc.) become more popular \cite{FiveDis}, the escalating complexity of their signal processing algorithms may preclude achieving low latency. Thus, new paradigms have to be introduced to address the above-mentioned challenges.

Deep learning (DL), mainly realized by deep neural networks (DNNs), has achieved impressive success with excellent results in diverse fields, such as image recognition \cite{Textbook}, mastering complex games like Go \cite{MasteringGame}, etc. This success has stimulated increasing interest in the application of DL in wireless communications. 
Specifically, DL is a powerful tool capable of learning the intricate inter-relationships of variables, especially those that are hard to accurately describe using mathematical models \cite{Textbook}. 
This enables us to design wireless communication systems without the knowledge of accurate mathematical models, which is impossible in the context of existing wireless design principles.
Moreover, for the family of light-weight DNNs having limited size, passing the inputs through them only requires a limited number of operations, which makes the DL methods computationally efficient. Even for highly complex DNNs which may be required for solving large-scale communication problems, the distributed parallel architectures and the accelerating tools of DL \cite{Textbook} are expected to result in a high computational efficiency. Thus, DNNs are attractive for solving large-scale wireless problems associated with numerous antennas/users/devices.

\begin{table*}
\centering
\caption{Fundamental differences between wireless transmission and DL}
\begin{tabular}{|l|c|c|}
 \hline
 & \tabincell{c}{Wireless transmission} & DL \\
\hline
Mathematical model & \tabincell{c}{Needs accurate mathematical model} & \tabincell{c}{Does not rely on accurate \\ mathematical model} \\
 \hline
 Design approach & \tabincell{c}{Optimizing each module separately \\ using mathematical derivation} & \tabincell{c}{Training the parameters \\ of the DNN as a whole} \\
 \hline 
Interpretation & Intuitive & Non-intuitive \\
\hline
\tabincell{c}{Generalization ability} & Widely applicable & Application-specific\\
\hline
Key challenges &  \tabincell{c}{Unrealistic assumptions} & \tabincell{c}{Too many parameters}  \\
\hline
\end{tabular}
\vspace{-3mm}
\label{difference}
\end{table*}

  Inspired by the advantages mentioned above, DL has been widely used for wireless communications. For instance, the state-of-the-arts in utilizing DL for physical layer communications are summarized in \cite{R1}. Moreover, \cite{R2} and \cite{R3} comprehensively survey the applications of DL in designing IoT and 5G cellular networks at various layers of the protocol stack, respectively. In contrast to the above reviews which eloquently survey the relevant literatures, we provide guidance on how to apply DL for wireless communications by inducing a pair of design
methodologies, namely DL-based architecture design and DL-based algorithm design. Particularly, DL-based architecture design utilizes DL to reformulate the traditional block-based communication design principle. This methodology is firstly exemplified by DL-based receiver design, followed by the more revolutionary DL-based joint transceiver design. Moreover, DL-based algorithm design utilizes DL to speed up the algorithmic processing at a guaranteed performance. To further explain it, DL-based transmission algorithm design and DL-based optimization algorithm design are illustrated by several 5G-style examples.
Their principles, key features, and performance gains will be discussed to shed light on these methodologies. More importantly, the intricate interplay between DL and wireless communications is highlighted.


The rest of the paper is organized as follows. Section II briefly summarizes the relationship between wireless communications and DL.  Sections III and IV detailedly investigate the intricacies of DL-based architecture design and  DL-based algorithm design, respectively.  Open problems and future research opportunities are provided in Section V, where the interplay between DL and wireless communications is further augmented. Finally, tangible conclusions and the further implications are offered in Section VI. 

\vspace{-1mm}

\section{Different Paradigms of Wireless Communications and DL}
Although there is an explosive proliferation of utilizing DL in wireless communications, DL is very different from wireless communications, especially from wireless transmission, which is summarized in Table \ref{difference}. Specifically, wireless transmission relies on accurate (but often simplified) mathematical models, such as the Additive White Gaussian Noise (AWGN) channel model, to design channel estimation algorithms or channel feedback schemes, etc. However, DL usually does not rely on such mathematical models of its tasks, and is particularly beneficial in the absence of the accurate mathematical models.
Moreover, traditional wireless transmission tends to design each module of the communication system separately using mathematical derivations, while DL usually trains all the parameters of the DNN as a whole. As a benefit of their mathematical models, classic methods of wireless transmission are often intuitive with plausible explanations and are widely applicable. By contrast, those of DL are sometimes non-intuitive, may even be hard to interpret, and application-specific, which means that different neural networks have to be trained for different tasks. Besides, wireless transmission often relies on idealized and simplified assumptions in their mathematical model, while DL may have an excessive number of parameters, making its training process time-consuming.  

Despite the significant differences between wireless transmission and DL, in this paper we focus on intrinsically amalgamating their benefit, while circumventing their weaknesses by portraying their profound interplay in the following sections. Specifically, in Section III we show that DL is capable of developing a new design paradigm for wireless transmission systems by introducing \cite{AnIntro}. Furthermore, in Section IV we demonstrate based on \cite{AMP} that the expert knowledge from wireless transmission can in turn help the development of DL techniques.


\begin{figure*}
\centering
\includegraphics[width=0.9\linewidth]{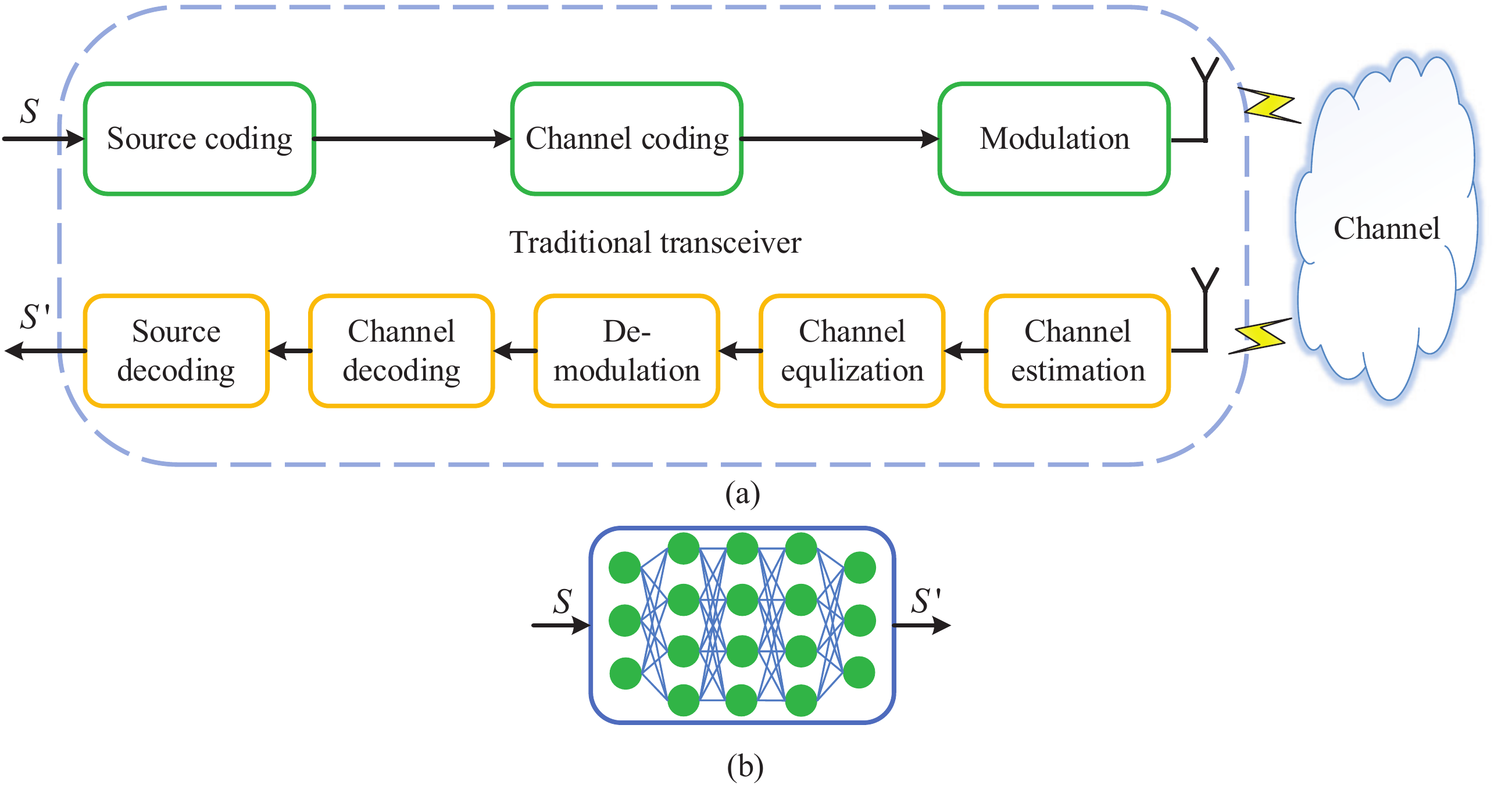}
\vspace*{-3mm}
\caption{From the traditional communication system to the DNN-based communication system: a) classic communication system consisting of many blocks; b) DNN-based communication system relying on one single block.}
\label{end_to_end}
\vspace*{-3mm}
\end{figure*}

\vspace{-2mm}
\section{DL-Based Architecture Design for Wireless Communications}
In this section, we focus on DL-based architecture design for intelligent wireless communication systems. 
Following the order spanning from designing the receiver to designing the whole system, 
we firstly introduce the DL-based receiver design, and then present the more revolutionary DL-based joint transceiver design. Particularly, DL-based receiver design invokes DL for jointly optimizing several
blocks of the receiver, which shows a beneficial performance gain in terms of the bit error rate (BER) when non-linear effects are encountered. By contrast, DL-based joint transceiver design optimizes the entire point-to-point communication system as
an end-to-end autoencoder, which results in a block error rate (BLER) that is lower than that of the 
traditional system using binary phase shift keying (BPSK).
 \vspace{-3mm}
\subsection{DL-Based Receiver Design}
For several decades, the block-based design principle has dominated wireless communication system design, where the communication system can be split into several independent functional blocks, including source coding, channel coding, etc., as shown in Fig. \ref{end_to_end} (a). Each block is optimized with the aid of mathematical model and expert knowledge, such as the optimal design of channel estimation which heavily relies on channel model and estimation theory.  This block-based structure provides convenience for system building, but it cannot cope with excessively complex scenario when channel model is unknown.

To deal with the unknown channel model when non-linear noise is introduced, a DL-based orthogonal frequency division multiplexing (OFDM) receiver is proposed in \cite{OFDM} to learn the channel behaviors and decode the signals. Moreover, channel estimation, channel equalization, and channel decoding are jointly designed in the receiver using DNN, which yields an architectural revolution for wireless communication systems. 
To be more specific, for OFDM systems suffering from a high peak-to-average-power-ratio (PAPR), peak-clipping and filtering
are usually used to alleviate the impact of PAPR. However, non-linear clipping noise will be introduced, which makes the equivalent channel more difficult to describe mathematically. Thus, mathematical model based channel estimation and equalization may become inaccurate, which imposes a certain performance loss during signal detection. To solve this problem, a fully connected DNN
is embedded into an OFDM receiver for signal detection in \cite{OFDM}, where the ReLU function is used as the activation function in the hidden
layers, and the sigmoid function is utilized in the last layer to map the result to the interval of [0,1]. In the training process, the original frequency-domain bit streams and their contaminated versions received after passing through the channel are used as labels and inputs, respectively. After that, the bit streams comprising the unknown frequency-domain signals are
fed into the trained DNN, and a threshold is applied to finally reconstruct the transmitted bits. Simulation results show that once the non-linear effects have been introduced by clipping, the DL-based OFDM receiver exhibits a beneficial performance gain, provided that the signal-to-noise-ratio (SNR) is in excess of 15 dB \cite{OFDM}.

 \begin{figure}
\centering
\includegraphics[width=0.9\linewidth]{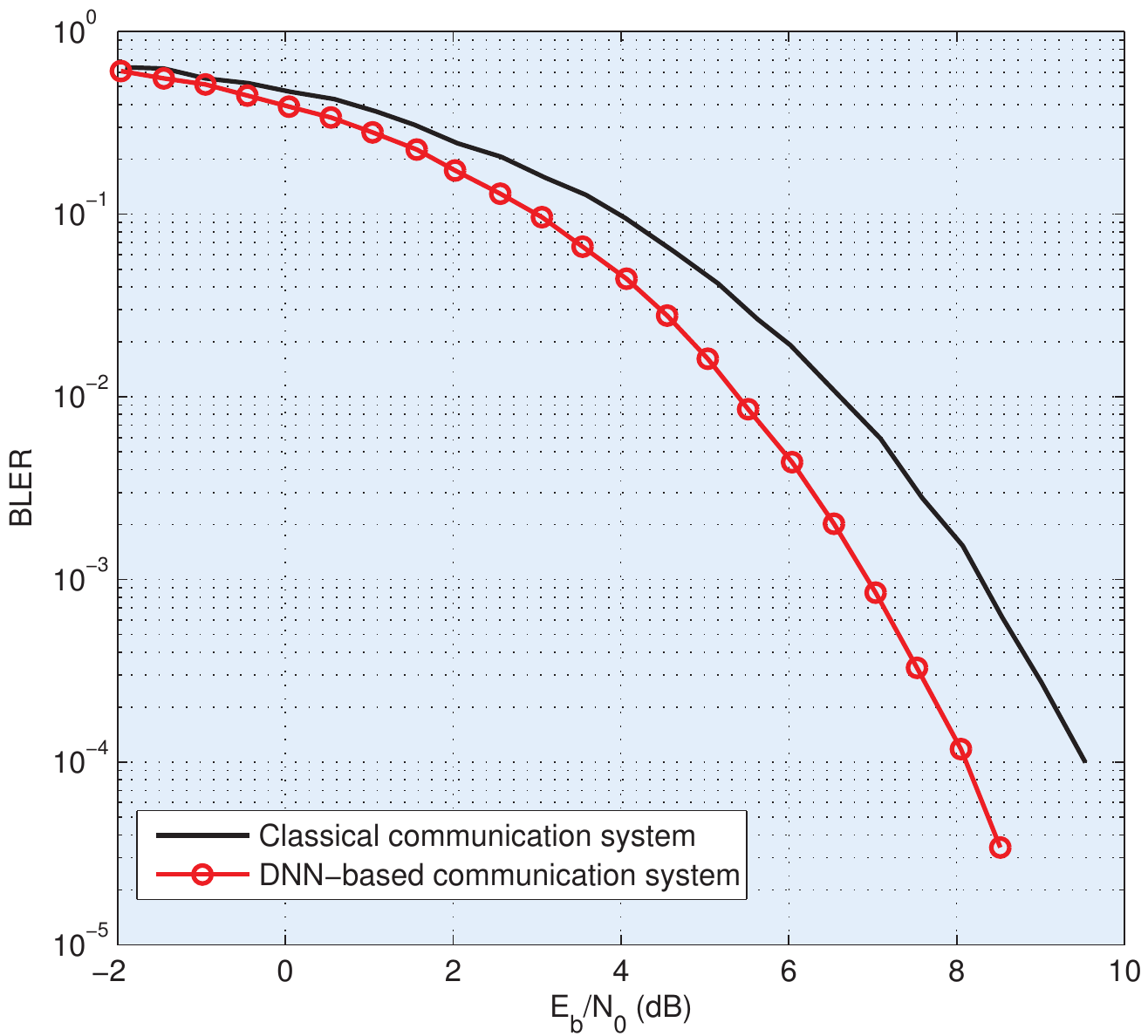}
\vspace*{-1em}
\caption{Block error rate (BLER) comparison between the block-based communication system and the DNN-based communication system.} 
\label{end_to_end_result}
\vspace*{-5mm}
\end{figure}

\begin{figure*}
\centering
\includegraphics[width=0.9\linewidth]{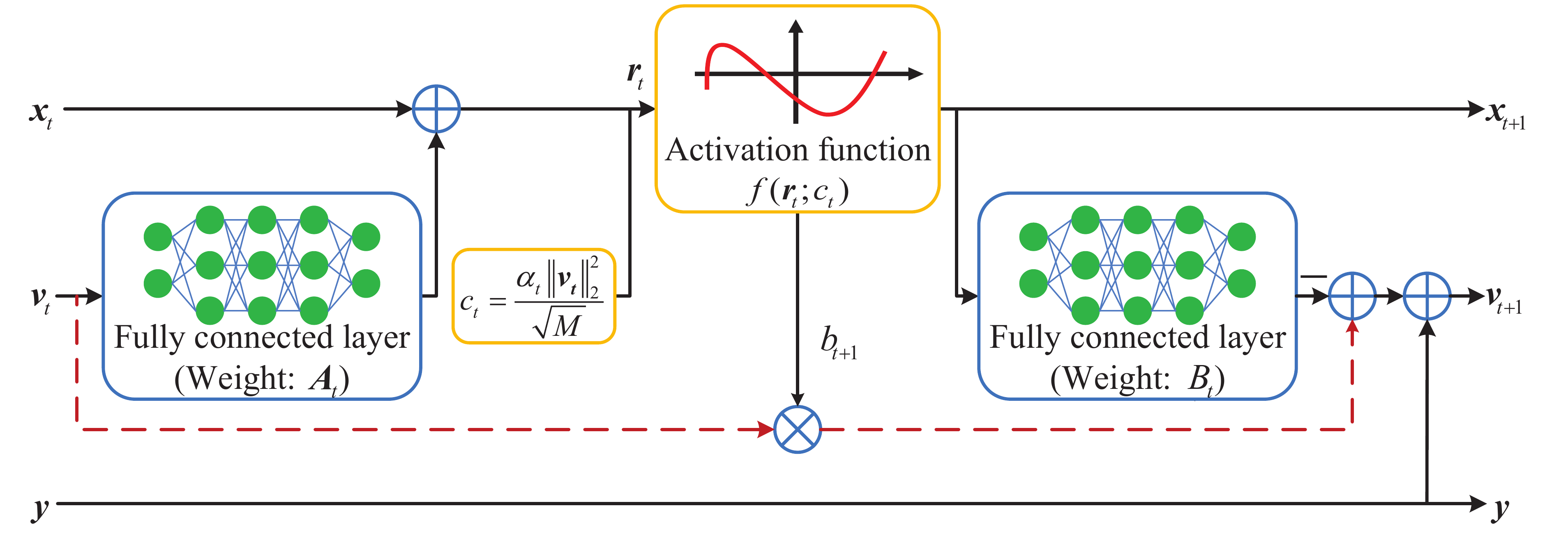}
\caption{The $t^{th}$ layer of the neural network utilized for DL-based AMP algorithm.
}
\label{AMP}
\vspace{-1em}
\end{figure*}
\vspace{-3mm}
\subsection{DL-Based Transceiver Design}
In contrast to the DL-based receiver design which only optimizes the receiver using DNN, a more revolutionary paradigm has been proposed recently to jointly design the whole point-to-point communication system (including both the transmitter and the receiver) as a single end-to-end autoencoder under the DL framework \cite{AnIntro}, which is completely different from the classic block-based design principle. This completely new paradigm of DNN-based communication system design is inspired by the resemblance of wireless communication systems and the autoencoder, both of which aim at equating the input message $s$ and the output message $s^\prime$. The DNN-based communication system is shown in Fig. \ref{end_to_end} (b), where the transmitted signal $s$ corresponds to one of the $k$-bit symbols from an alphabet, which is encoded as an $M$-dimensional one-hot vector ($M=2^k$) and used as the input of the autoencoder. After passing through the transmitter composed of a series of hidden layers and a normalization layer for meeting the energy constraint, an $n$-dimensional vector $\bm{x}$ is generated as the encoded signal. Then, a Gaussian noise layer serves as the AWGN channel to contaminate the original signal $\bm{x}$ to become $\bm{y}$, which will be fed into the receiver having multiple hidden layers and terminated by a softmax layer to output the reconstruction probabilities of all possible symbols. Finally, the symbol associated with the highest probability will be chosen as the recovered signal $s^\prime$. 
As illustrated in Fig. \ref{end_to_end_result}, simulation results generated using Tensorflow \cite{Textbook} show that
the DNN-based communication system outperforms the traditional system using BPSK in terms of its BLER \cite{AnIntro}, because the new scheme is capable of jointly optimizing the communication system as a whole rather than optimizing individual blocks in Fig. \ref{end_to_end} (a) in isolation. 
This new design paradigm allows us to possibly replace all baseband components of the block-based communication system of Fig. \ref{end_to_end} (a) by a single DNN of Fig. \ref{end_to_end} (b) , which can be trained by exploiting the seemingly infinite computation and storage resources of the cloud. 

\vspace*{-1mm}
\section{DL-Based Algorithm Design for Wireless Communications}

Due to its convenient implementation relying on parallel architectures and the powerful learning capability, DL can also be used to speed up the algorithmic processing at a potentially improved performance and reduced latency. In this section, we will introduce the methodology of DL-based algorithm design for wireless communications, which will be illustrated from two aspects, i.e., DL-based transmission algorithm design and DL-based optimization algorithm design, using several typical techniques conceived for 5G and beyond, including mmWave massive MIMO, LDPC, NOMA, and UDN. 

\setcounter{subsection}{0}
\vspace*{-3mm}
\subsection{DL-Based Transmission Algorithm Design}

Since reliable transmission is one of the most important tasks in wireless communications, we will introduce DL-based transmission algorithm design in this part. Following the order of signal processing, i.e., steps commencing from the inner receiver to the outer receiver, channel estimation is carried out before channel decoding.
Therefore, we investigate DL-based sparse channel estimation for mmWave massive MIMO first, and then introduce DL-based belief propagation for LDPC decoding, both of which constitute typical 5G-style techniques.


\subsubsection{{DL-Based Sparse Channel Estimation for MmWave Massive MIMO}}
MmWave massive MIMO is one of the most promising techniques in 5G conceived for largely increasing the data rates \cite{MIMO,Xiuhong}. It requires accurate CSI to realize multi-user precoding, decoding, etc. Since the mmWave channel is sparse in the angular domain \cite{MIMO,Xiuhong}, compressive sensing algorithms have been widely used for sparse mmWave channel estimation, such as approximate message passing (AMP). However, AMP requires many iterations of linear residual updating and non-linear shrinkage operation to converge, which makes it difficult to
guarantee both reliable performance and reduced latency at the same
time. 

To address this issue, a DL-based AMP algorithm has been proposed in \cite{AMP} to significantly reduce the number of iterations required under a specific performance constraint. As shown in Fig. \ref{AMP}, each iteration of the AMP algorithm is unfolded to a single layer of the DNN, where the linear residual updating is accurately modeled by the linear operations of the DNN, and the non-linear shrinkage operation is realized by the activation function in the DNN. 
Additionally, unlike the classical AMP algorithm which fixes the parameters of the shrinkage operation in advance, the DL-based AMP actually learns those parameters during the training process. It has been shown by simulation results that the DL-based AMP converges faster and outperforms both the AMP and the iterative shrinkage/threshold algorithm (ISTA) in terms of its average normalized mean square error (NMSE). The improved performance is partly attributed to the introduction of the Onsager correction into AMP \cite{AMP}, and partly contributed by the DNN-aided learning of optimal shrinkage parameters. 
Thus, expert knowledge gleaned from other fields may in turn help to enhance the DL performance. More importantly, the authors of \cite{AMP} revealed basic principles of designing new DNN architectures suitable for wireless communications by unfolding the iterations of the existing algorithms into layers of a DNN. This is especially useful when the complexity of the existing iterative algorithms is excessive for practical implementation of massive communications with large-scale signal processing.

\subsubsection{DL-Based Belief Propagation for LDPC Decoding}
In 2017, low-density parity-check (LDPC) coding has been chosen to replace the 4G turbo coding as the new channel coding scheme in the most important broadband mode of the 5G standard. This can be attributed to the excellent performance of LDPC coding operating close to the Shannon limit in AWGN channels. However, due to the detection effect of filtering, oversampling, and multi-user interference in practical systems \cite{BPCNN}, the received signals may become contaminated by colored noise, which is hard to mathematically model and would impose an obvious performance loss on the LDPC decoding performance.

To solve this problem, Liang \textit{et al.} \cite{BPCNN} modified the widely used belief propagation (BP) algorithm by cascading a Convolutional Neural Network (CNN) to the standard BP decoder, which succeeds in outputting a more accurate noise estimate from the decoding result of the BP decoder. In the BP-CNN decoder, the received signal $y$ will firstly be fed into the standard BP decoder to obtain an initial estimate of the transmitted signal $\hat{x}$, which will then be subtracted from $y$ to obtain the estimated noise $\hat{n}=y-\hat{x}$. The following trained CNN exploits this estimated noise $\hat{n}$ and outputs a more accurate noise estimate $\tilde{n}$, which will be fed back to the BP decoder and then subtracted from the received signal $y$ to obtain a more ``pure signal'' $\tilde{y}$. A new round of operations will be applied to the newly generated $\tilde{y}$, and iterating between BP and CNN as mentioned above can finally achieve an improved LDPC decoding performance, which mitigates the adverse effects imported by the colored noise.

As for the simulation results, the total number of BP iterations is set to 50 in \cite{BPCNN} for both the standard BP and the improved BP-CNN, and the latter outperforms the former in terms of its BER. Therefore, to achieve the same BER performance, the BP-CNN requires a much lower number of iterations than the standard BP, indicating a reduced latency for channel decoding. Moreover, since the improved BP-CNN does not rely on any specific channel coding scheme or channel model, its design principle may also be applied to other codes, such as turbo codes and BCH codes, which suggests that the physical layer can be quite flexibly designed.   


\vspace{-3mm}
\subsection{DL-Based Optimization Algorithm Design}
Optimization plays an important role in wireless communication systems to realize efficient exploitation of limited radio resources. However, many optimization algorithms require a large number of iterations to converge, which results in both high complexity and high latency, especially when the problem's scale is very large.
In this part, we will show that DL can be used to speed up processing while maintaining reliable performance by introducing two works. Particularly, following the order spanning from deterministic networks to random networks, DL-based sum rate maximization for NOMA with predetermined transceivers is introduced first, followed by the introduction of the DL-based energy consumption minimization for UDN, where the transceivers are randomly paired according to the distance, link quality, etc. The two above-mentioned solutions also constitute typical 5G technologies.

\subsubsection{{DL-Based Sum Rate Maximization for NOMA}}
Power allocation plays an essential role in the emerging 5G NOMA solutions \cite{NOMA}, and it has attracted extensive attention in recent years. To solve this problem, a series of optimization algorithms have been proposed, among which the weighted minimum mean square error (WMMSE) is quite popular \cite{NOMA}. However, since optimization algorithms like WMMSE usually require a large number of iterations to converge, it is difficult for them to achieve the requirement of low-latency. 

To deal with this challenge, a DL-based optimization approach is proposed in \cite{LearningTo} to accelerate processing, while maintaining reliable performance in an interference-contaminated wireless scenario supporting $K$ single-antenna transceiver pairs. Furthermore, the power of each pair of the transceivers is limited by a certain constraint, and accordingly a sum rate optimization problem is formulated under this constraint.
To solve this optimization problem faster, Sun \textit{et al.} \cite{LearningTo} interpreted the complex WMMSE algorithm as an unknown non-linear mapping between its inputs (system parameters) and outputs (power allocation results), and used a DNN having multiple hidden layers to mimic the WMMSE algorithm as accurately as possible, exploiting its high computational efficiency in its testing process for promptly finding a beneficial power allocation solution. Furthermore, the ReLU function \cite{Textbook} is used as the activation function, and the MMSE is calculated as the loss function. The channel coefficients obeying the classic Rayleigh distribution and the power
allocation results generated by WMMSE are adopted as data and labels to train the DNN. Moreover, the simulation time comparison of the standard WMMSE algorithm and the DL-based optimization method is presented in \cite{LearningTo}. All simulation codes of these two methods are written in Python and run on the same computer. It is observed that the DL-based method is capable of reducing the execution time by a factor of several hundreds compared to the standard WMMSE algorithm, while maintaining a similar sum rate performance. For example, 97.92\% of the max sum rate can be achieved when supporting 10 users. Moreover, as the number of users increases from 10 to 30, the execution time of the standard WMMSE method increased from 12.32s to 78.06s, while that of the DL-based method grows steadily from 0.04s to 0.09s, which shows a modest complexity increase of the DL-based method for large-scale problems.


\subsubsection{{DL-Based Energy Consumption Minimization for UDN}}
UDN is another promising technology for 5G communication systems, where a large number of access points (APs) are used to provide high data rates. Each AP can set up a transmission link to another one within its transmission range, where minimizing the energy consumption of dozens of APs while satisfying the flow demand is very important.  

In order to cope with the associated large-scale optimization problem, diverse methods have been proposed to reduce the complexity, but very few approaches have been designed for directly reducing the problem scale. To this end, a DL-based algorithm is proposed in \cite{Optimization} to reduce the
 problem scale, while the optimality of the solution can be maintained at the same time. Since there are always a number of transmission links with no flow passing through in the optimal case, a DNN with multiple hidden layers is utilized to predict those ``deactivated links'' and eliminate them from the original large-scale optimization problem, thus directly reducing the problem scale. To be more specific, the flow demand vectors are used as input, and the flows of the links calculated by the so-called delay column generation (DCG) \cite{Optimization} applied to the original optimization problem are used as labels to train the DNN. In the testing stage, after obtaining the values of flows passing through the links, a threshold is applied to eliminate ``deactivated links'' in preparation for the problem reformulation. Finally, the DCG will then be used again for the scale-reduced optimization problem to obtain the final results. The simulation results of \cite{Optimization} show that the prediction is quite accurate, and the optimization problem can be solved within at most half of
the original time, while the performance difference 
between the original large-scale problem and the scale-reduced one is not larger than 3\%.

It should be pointed out that the idea of the elimination-based method advocated in \cite{Optimization} does not rely on any specific scenario or algorithm, hence it can also be extended to reduce the scale of other optimization problems by excluding zero variables.

\vspace*{-2mm}
\section{Challenges and Research Opportunities}
Although DL-based methods proposed for wireless communications in the previous two sections have shown encouraging advantages over their classic counterparts, they are still in their infancy, and there are many challenging issues remaining for further study. In this section, we underline the important challenges in this emerging area from four perspectives, namely theoretical challenges, data-related challenges, algorithmic challenges, and implementation-oriented challenges. The corresponding potential solutions and research opportunities are also highlighted. Furthermore, we believe that the study of DL-based wireless communications can also help to promote the development of DL itself by addressing these challenges. 

\setcounter{subsection}{0}
\vspace*{-3mm}
\subsection{Theoretical Challenges}
\subsubsection{Performance Analysis}
In contrast to the classical methods of wireless communications under the umbrella of Shannon's information theory, DL-based methods lack solid mathematical foundations in terms of theoretical analysis, which is however desired to guide the practical design of DL-based wireless communications and provide insights into their performance limits.
Provided that the training data obeys a certain distribution, Shannon information theory may still contribute to the theoretical basis of DL. But again, this is an open problem requiring further exploration.

\subsubsection{Interpretability}
 The mathematical or physical interpretation of DL-based methods is in its infancy. For instance, the selection of training strategies remains somewhat haphazard without clear physical explanation, and we cannot easily make plausible why a certain network structure attains a good performance. Since the training process itself is actually an optimization problem, advanced optimization theory is expected to help us find the most appropriate loss function and training strategy. Moreover, as indicated by the DL-based AMP algorithm of \cite{AMP}, the expert knowledge gleaned from other fields (such as wireless communications) may in turn help the design of more efficient DNN architectures for wireless communications. 

\vspace*{-3mm}
\subsection{Data-Related Challenges}
The amount and quality of training data are essential for the final performance of DL.
However, acquiring sufficiently high-quality data for training in real wireless communication systems is not as straightforward as in data-oriented applications, which is mainly due to the following two reasons. Firstly, how to generate a large amount of data remains a challenge. For example, transmitting a large packet of pilots as labeled data is very difficult in wireless environments due to high spectrum efficiency requirement. Secondly, data in wireless communications are usually high-dimensional (e.g., massive MIMO channels) and highly heterogeneous (since diverse services associated with significantly different quality of service requirements have to be supported). Hence, pre-processing and cleaning of the available data will be challenging. Thus, a mature data processing flow should be developed from data collection to pre-processing. Moreover, as DL is known to be inherently data-driven, different data sets used for training and testing may result in different performance. To ensure fair comparison, a promising solution is to build some widely accepted and low-cost data sets to test different DL-based algorithms for wireless communications. 

\vspace*{-3mm}
\subsection{Algorithmic Challenges}
Communication scenarios are rather complicated and usually time-varying. Thus, the data collected may not be able to cover all situations, or to reflect all changes in a timely manner. 
For instance, DL is not directly suitable for dealing with time-varying environments. To be more specific, if the wireless channel changes rapidly, the DL-based wireless systems may have to be frequently and completely re-trained from scratch to maintain their performance over time, which is time consuming and computationally expensive. Thus, DNN-based methods have to be robust and versatile in similar but hitherto uncovered situations, and new learning algorithms should be designed. For example, transfer learning \cite{Textbook} may be a possible solution, which enable DNN to learn from similar cases so as to be also applicable to new scenarios, thus achieving better environmental adaptability and robustness. We believe that the study of DL for such scenarios may also promote its development in terms of both theory and algorithmic innovation.

\vspace{-3mm}
\subsection{Implementation-Oriented Challenges}
Apart from the challenges listed above, deploying DL-based methods in real communication systems will face implementation-oriented challenges. Firstly, most of the communication infrastructure, such as the base stations, are only equipped with radio frequency (RF) and signal processing functionality. To implement DL-based methods, a cloud server should be deployed based on the existing infrastructure. Secondly, since the classic communication methodologies may not be completely replaced by the DL-based techniques in a short time, a soft switching scheme is needed to switch between the DL-based methods and the non DL-based methods. 
We may need a new architecture to support the harmonious co-existence of DL-based methods and the classic non DL-based methods. 

\vspace{-1mm}
\section{Conclusions}

In this review, we have revealed a pair of
dominant methodologies for the applications of DL in wireless communications, namely DL-based
architecture design and DL-based algorithm design. We have also analyzed their design principles, key features, as well as performance gains, and demonstrated that DL is indeed capable of supporting communication systems in complex operating scenarios and speeding up large-scale processing with guaranteed performance. Moreover, we have pointed out some challenges and research opportunities in this emerging area. In summary, we believe that further study of this area can also help to promote the
development of DL itself in theory and algorithmic innovation. 


%


\vspace{-9mm}
\begin{IEEEbiography}[{\includegraphics[width=0.9in,height=1.1in,keepaspectratio]{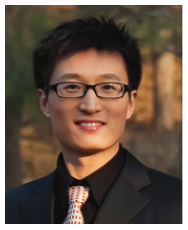}}]{Linglong Dai}
(M’11, SM’14) received the Ph.D. degree from Tsinghua University, Beijing, China, in 2011. He is currently an associate professor at Tsinghua University. His current research interests include massive MIMO, millimeter-wave/THz communications, reconfigurable intelligent surface, multiple access, and sparse signal processing. He has received five IEEE conference best paper awards, the Electronics Letters Best Paper Award in 2016, the National Natural Science Foundation of China for Outstanding Young Scholars in 2017, the IEEE ComSoc Asia-Pacific Outstanding Young Researcher Award in 2017, the 7th IEEE ComSoc Asia-Pacific Outstanding Paper Award in 2018, the China Communications Best Paper Award in 2019, and the IEEE Communications Society Leonard G. Abraham Prize in 2020.
\end{IEEEbiography}

\begin{IEEEbiography}[{\includegraphics[width=0.9in,height=1.1in,keepaspectratio]{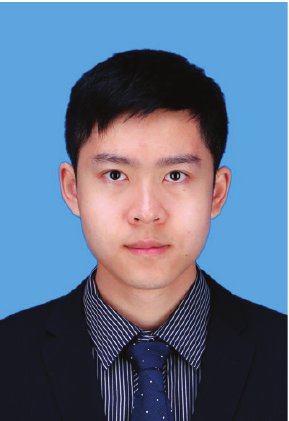}}]{Ruicheng Jiao}
(S'19) received his B.S. degree in Physics from
Tsinghua University, Beijing, China, in 2016. He is currently
working towards the Ph.D. degree in the Department of Electronic
Engineering, Tsinghua University, Beijing, China. His research
interests are in mmWave communications, new multiple
access techniques, and deep learning for future wireless communications.
\end{IEEEbiography}

\begin{IEEEbiography}[{\includegraphics[width=0.9in,height=1.1in,keepaspectratio]{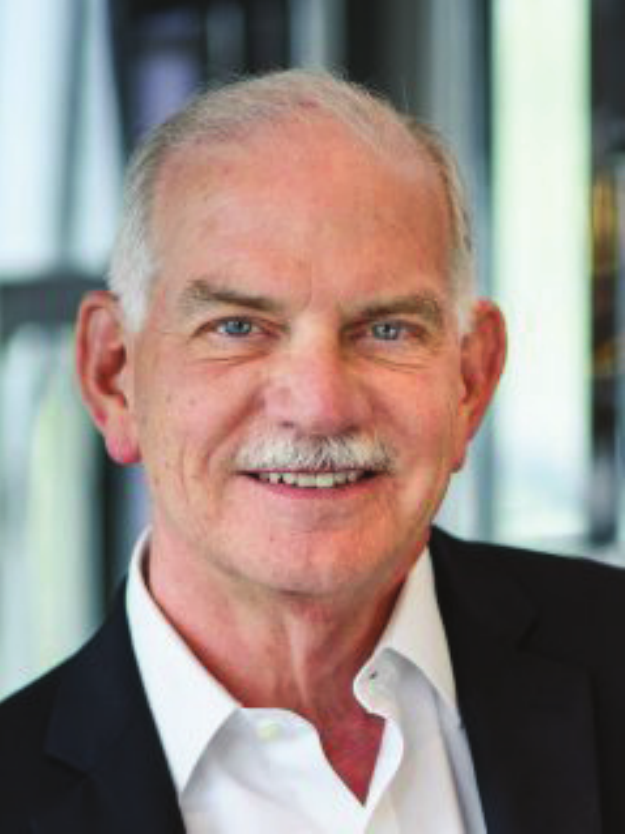}}]{H. Vincent Poor}
(S'72-M'77-SM'82-F'87) is the Michael Henry Strater University Professor at Princeton University. His interests lie in the areas of information theory, machine learning and network science, and their applications in wireless networks, energy systems and related fields. A member of the U.S. National Academies of Engineering and Sciences, he has received the IEEE Communication Society’s Marconi (2008), Armstrong (2009), Hertz (2018) and Ellersick (2020) Awards, as well as the IEEE Bell Medal (2017).
\end{IEEEbiography}

\begin{IEEEbiography}[{\includegraphics[width=0.9in,height=1.1in,keepaspectratio]{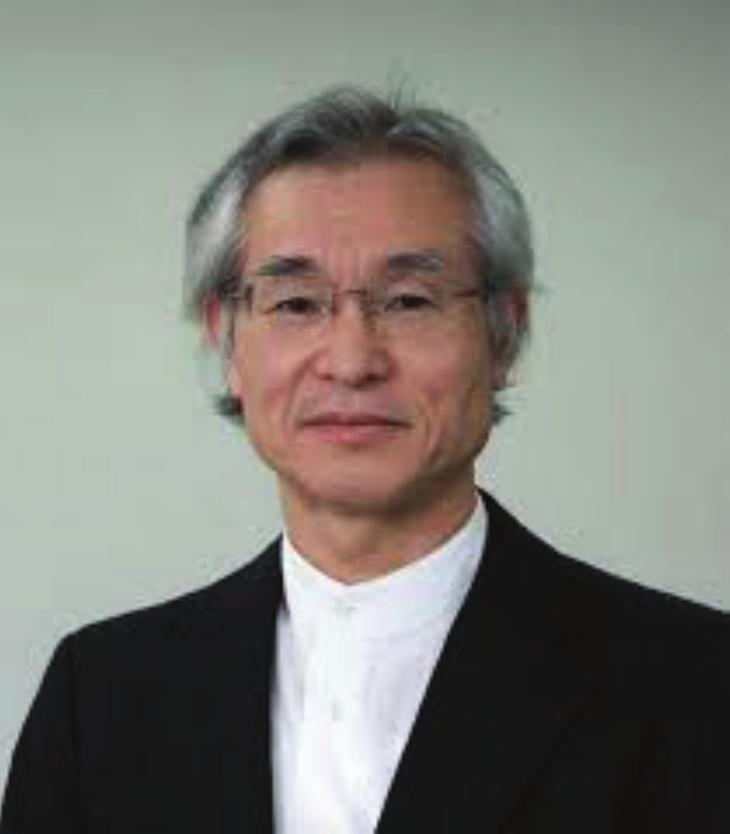}}]{Fumiyuki Adachi}                 (IEEE LF’16, IEICE F'07) received the B.S. and Dr. Eng.
degrees in electrical engineering from Tohoku University, Sendai, Japan,
in 1973 and 1984, respectively. In April 1973, he joined the Electrical
Communications Laboratories of Nippon Telegraph \& Telephone Corporation
(now NTT) and conducted research on digital cellular mobile
communications. From July 1992 to December 1999, he was with NTT DOCOMO,
where he led a research group on wideband/broadband wireless access for
3G and beyond. He contributed to the development of 3G air interface
standard, known as W-CDMA. Since January 2000, he has been with Tohoku
University, Sendai, Japan. He was a full professor at the Dept. of
Communications Engineering of the Graduate School of Engineering until
he retired from the university in March 2016. Currently, he is a
Specially Appointed Professor for Research at the Research Organization
of Electrical Communication (ROEC), Tohoku University and is leading a
wireless signal processing research group aiming at beyond 5G systems.
\end{IEEEbiography}

\begin{IEEEbiography}[{\includegraphics[width=0.9in,height=1.1in,keepaspectratio]{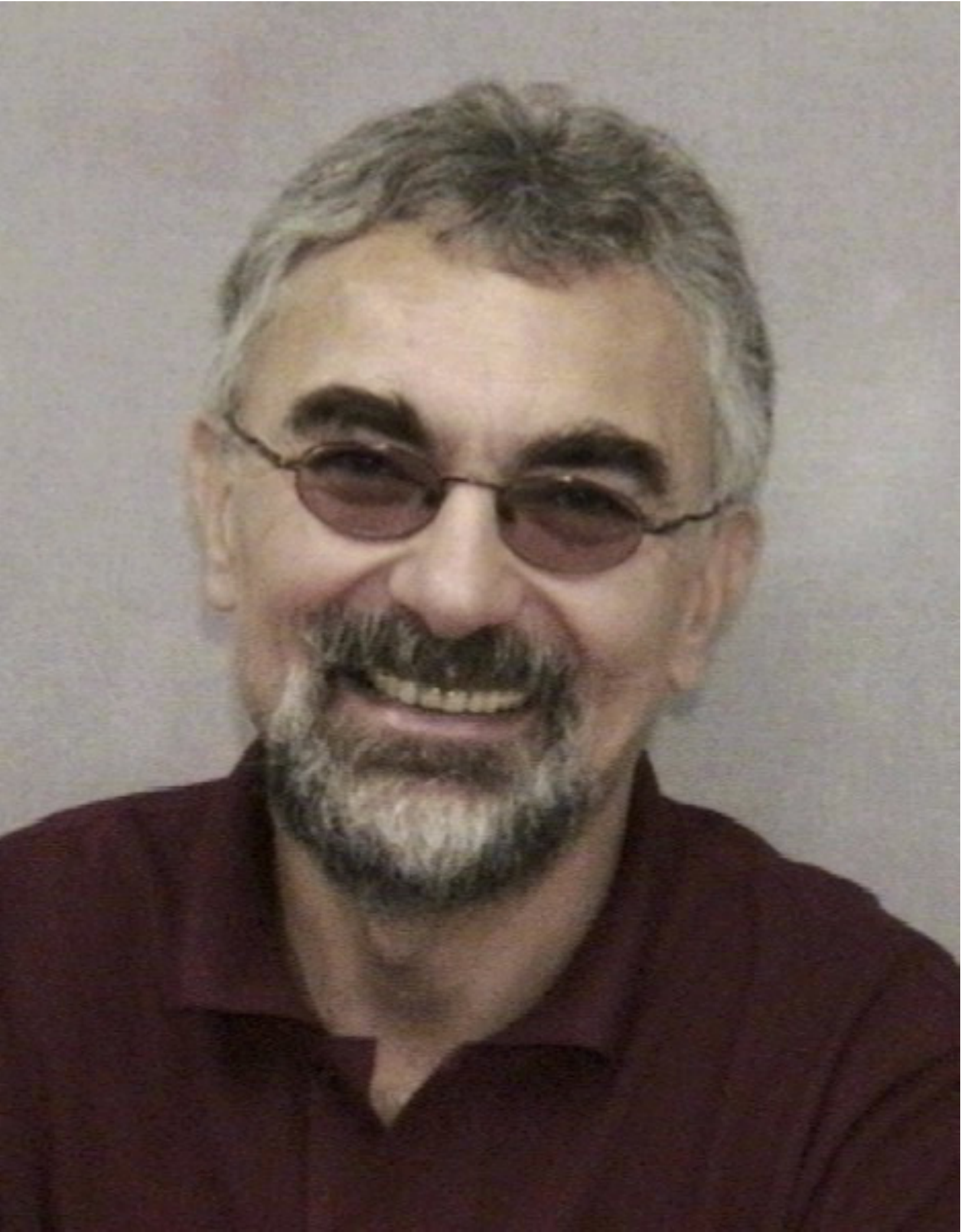}}]{Lajos Hanzo}
(https://www-mobile.ecs.soton.ac.uk, https://en.wikipedia.org/wiki/Lajos\_Hanzo), FIEEE'04, Fellow of the
Royal Academy of Engineering F(REng), of the IET and of EURASIP,
received his Master degree and Doctorate in 1976 and 1983,
respectively from the Technical University (TU) of Budapest. He was
also awarded the Doctor of Sciences (DSc) degree by the University of
Southampton (2004) and Honorary Doctorates by the TU of Budapest
(2009) and by the University of Edinburgh (2015).  He is a Foreign
Member of the Hungarian Academy of Sciences and a former
Editor-in-Chief of the IEEE Press.  He has served several terms as
Governor of both IEEE ComSoc and of VTS.  He has published 1900+
contributions at IEEE Xplore, 19 Wiley-IEEE Press books and has helped
the fast-track career of 123 PhD students. Over 40 of them are
Professors at various stages of their careers in academia and many of
them are leading scientists in the wireless industry.
\end{IEEEbiography}








\ifCLASSOPTIONcaptionsoff
  \newpage
\fi

\end{document}